\documentclass[longauth]{aa}
\usepackage{txfonts,graphicx,natbib}
\bibpunct{(}{)}{;}{a}{}{,}

\begin{document}

\title{MAGIC observation of the GRB\,080430 afterglow}

%
\author{
 J.~Aleksi\'c\inst{1} \and
 H.~Anderhub\inst{2} \and
 L.~A.~Antonelli\inst{3} \and
 P.~Antoranz\inst{4} \and
 M.~Backes\inst{5} \and
 C.~Baixeras\inst{6} \and
 S.~Balestra\inst{4} \and
 J.~A.~Barrio\inst{4} \and
 D.~Bastieri\inst{7} \and
 J.~Becerra Gonz\'alez\inst{8}
% ...
}
\institute { IFAE, Edifici Cn., Campus UAB, E-08193 Bellaterra, Spain
 \and ETH Zurich, CH-8093 Switzerland
 \and INAF National Institute for Astrophysics, I-00136 Rome, Italy
 \and Universidad Complutense, E-28040 Madrid, Spain
 \and Technische Universit\"at Dortmund, D-44221 Dortmund, Germany
 \and Universitat Aut\`onoma de Barcelona, E-08193 Bellaterra, Spain
 \and Universit\`a di Padova and INFN, I-35131 Padova, Italy
 \and Inst. de Astrof\'{\i}sica de Canarias, E-38200 La Laguna, Tenerife, Spain
% ...
}

\date{Received <date> / Accepted <date>}

\abstract {Gamma-ray bursts are cosmological sources emitting radiation from the gamma-rays to the radio band. Substantial observational efforts have been devoted to the study of gamma-ray bursts during the prompt phase, i.e. the initial burst of high-energy radiation, and during the long-lasting afterglows. In spite of many successes in interpreting these phenomena, there are still several open key questions about the fundamental emission processes, their energetics and the environment.} 
{Independently of specific gamma-ray burst theoretical recipes, spectra in the GeV/TeV range are predicted to be remarkably simple, being satisfactorily modeled with power-laws, and therefore offer a very valuable tool to probe the extragalactic background light distribution. Furthermore, the simple detection of a component at very-high energies, i.e. at $\sim 100$\,GeV, would solve the ambiguity about the importance of various possible emission processes, which provide barely distinguishable scenarios at lower energies.}
{We used the results of the MAGIC telescope observation of the moderate resdhift ($z\sim0.76$) \object{GRB\,080430} at energies above about 80\,GeV, to evaluate the perspective for  late-afterglow observations with ground based GeV/TeV telescopes.} {We obtained an upper limit of $F_{\rm 95\%\,CL} = 5.5 \times 10^{-11}$\,erg\,cm$^{-2}$\,s$^{-1}$ for the very-high energy emission of \object{GRB\,080430}, which cannot set further constraints on the theoretical scenarios proposed for this object also due to the difficulties in modeling the low-energy afterglow. Nonetheless, our observations show that Cherenkov telescopes have already reached the required sensitivity to detect the GeV/TeV emission of GRBs at moderate redshift ($z \lesssim 0.8$), provided the observations are carried out at early times, close to the onset of their afterglow phase.} {}


\maketitle

\authorrunning{Anderhub et al. (MAGIC collaboration)}

\section{Introduction}
\object{GRB\,080430} was detected by the \textit{Swift} satellite \citep{Geh04} on April 30, 2008 at 19:53:02\,UT \citep{Gui08}. The prompt emission lasted $\sim 16$\,s \citep{Stam08} allowing to assign this event to the long duration class \citep{Kouv93}. X-ray and optical counterparts were discovered and followed-up by many groups. Optical spectroscopy was rapidly carried out allowing to derive a redshift of $z = 0.758$. The redshift estimate has been revised recently with a more accurate wavelength calibration \citep[][and de Ugarte Postigo et al. in preparation, hereinafter DEUG10]{deUg08,CuFo08}. The relatively modest redshift made it an interesting target for the Major Atmospheric Gamma-ray Imaging Cherenkov (MAGIC) telescope\footnote{http://wwwmagic.mpp.mpg.de/} observations. In the past, upper limits for several Gamma-Ray Bursts (GRBs) at energies greater than about 100\,GeV were reported both for single event observations and for a sample of events \citep[e.g.][]{Alb06,Tam06,Alb07,Aha09}. In this paper we try to predict the Very-High Energy (VHE) flux for \object{GRB\,080430} by modeling the detected X-ray and optical afterglow and adopting as a reference the cosmological fireball model \citep{Pir99,Zha07}. In Sect.\,\ref{sec:magicobs} we report the results of the MAGIC observation, in Sect.\,\ref{sec:aft} we discuss the lower energy afterglow, in Sect.\,\ref{sec:ssc} we introduce the adopted modeling scenario for the VHE flux, in Sect.\,\ref{sec:ebl} we discuss the effect of Extragalactic Background Light (EBL) attenuation and finally, in Sect.\,\ref{sec:concl}, conclusions and considerations about future perspectives are drawn. Throughout the paper we assume a $\Lambda{\rm CDM}$ cosmology with $\Omega_{\rm m} = 0.27$, $\Omega_\Lambda = 0.73$ and $h_0 = 0.71$. At the redshift of the GRB the luminosity distance is $\sim 4.8$\,Gpc ($\sim 1.5 \times 10^{28}$\,cm, corresponding to a distance modulus $\mu = 43.4$\,mag). All errors are $1\sigma$ unless stated otherwise. Throughout this paper the convention $Q_x = Q/10^x$ has been adopted in CGS units. Results presented in this paper supersede those reported in \citet{Cov09b}.

\section{MAGIC observations}
\label{sec:magicobs}

\object{GRB\,080430} occurred while the Sun was still above the horizon at the MAGIC site (Roque de los Muchachos, $28.75^\circ$N, $17.89^\circ$W). The MAGIC observation started immediately after sunset, at 21:12:14\,UTC and ended at 23:52:30\,UTC. The observation 
was disturbed by clouds. The beginning of the observation was at T$_0$ + 4753\,s, well after the end of the prompt emission phase. The observation with MAGIC started at a zenith angle of Z$_{\rm d} = 23^\circ$, reaching Z$_{\rm d} = 35^\circ$ at the end. The data set was divided into two time intervals. Results from the first time interval, giving the lower energy upper limit, are used in this context. Analysis of the dataset, in the energy bin from 80 up to 125\,GeV with the spectral parameters derived in Sect.\,\ref{sec:ssc}, gave a 95\% CL upper limit of $F_{\rm 95\%\,CL} = 5.5 \times 10^{-11}$\,erg\,cm$^{-2}$\,s$^{-1}$ (under the assumption of steady emission) or a fluence limit of F$_{\rm 95\%\,CL} = 3.5 \times 10^{-7}$\,erg\,cm$^{-2}$ for a time interval of 6258\,s from 21:12:14 to 22:56:32\,UTC. These limits contain a 30\% systematic uncertainty on the absolute detector efficiency. Limits at higher energies are less important for the present analysis due to intense EBL absorption above $\sim 100$\,GeV (see Sect.\,\ref{sec:ebl}). It is important to note that at that time, the sum trigger hardware upgrade \citep{Alb08,GRB} which allows the MAGIC telescope to carry out reliable observations with lower energy threshold was not yet available for GRB observations. Therefore the lowest obtained upper limit is a factor two higher than in later cases \citep[e.g.][]{GRB090102}.

\section{Afterglow light-curve and spectral energy distribution}
\label{sec:aft}

It is not our purpose to discuss here the physics of the afterglow of this event, which will be discussed in detail in DEUG10. Nevertheless, preliminary results shows that this afterglow can not be satisfactory described within any common referred scenario. In particular the early afterglow is puzzling, likely requiring two distinct components with separated time evolution. However, at the epoch of the MAGIC observations (about 8\,ksec from the high-energy event), the afterglow seems to have entered a more stable phase although other components, as late prompt emission, can still be contributing \citep{Ghi07}. Analysis of the spectral (from optical to X-rays) information shows that the afterglow can be described as due to the interaction of a relativistic outflow with the circumburst medium surrounding the progenitor \citep{Pir99,Zha07}. The outflow is relativistic and shocks form with consequent particle acceleration. Details of the acceleration process are not known and it is usually assumed that electrons follow a power-law distribution in energy with a slope $p$. Numerical simulations suggest it should be $p \sim 2.2-2.3$ \citep{Acht01,Vie03} although other scenarios predict a wider range which is indeed supported by the analysis of several afterglows \citep{ElDo04,Shen06}. The late-afterglow of GRB\,080430 can be characterized by a constant circumburst density environment with typical number density $n \sim 1$\,cm$^{-3}$. The electron distribution index turns out to be $p \sim 2.1$. Given the afterglow spectral properties, it is possible to predict the time decay, which in the optical, is well consistent with the predictions. On the contrary, X-ray data \citep{Gui08b} show a much milder decay than expected. It is difficult to attribute this behaviour to a specific physical ingredients. Common additions to the reference model \citep{Zha07}, which may or may not modify the spectrum involve late energy injection, structured jets, flares, circumburst density variations, etc. \citep[see e.g.][for comprehensive discussions about these factors]{Pan06,Zha06}. It is clearly well beyond the scope of this paper to discuss these issues in detail, which are indeed still not well settled. We therefore model the VHE emission assuming the afterglow could be described in the context of the standard afterglow model \citep{Pir99,Zha07}. Finally, we comment possible modifications induced by additional phenomena which in general can even increase the expected VHE flux.

In order to characterize the afterglow spectrum we must compute the synchrotron injection, $\nu_{\rm m}$, and cooling, $\nu_{\rm c}$, frequency values. The injection frequency is where most of the synchrotron emission occurs and the cooling frequency identifies where electrons cool effectively. In case of constant circumburst medium \citep{Yos03,FaPi06} we have:

\begin{equation}
\nu_{\rm m} \approx 4.3 \times 10^{14}~C_p^2\,\epsilon_{{\rm B},-2}^{1/2}\,\epsilon_{{\rm e},-1}^{2} E_{\rm k,53}^{1/2}\,t_3^{-3/2}\,(1+z)^{1/2}~\,{\rm Hz},
\label{eq:v}
\end{equation}
and
\begin{equation}
\nu_{\rm c} \approx 1.8 \times 10^{16}~\epsilon_{{\rm B},-2}^{-3/2}\,n^{-1} E_{\rm k,53}^{-1/2}\,t_3^{-1/2}\,(1+z)^{-1/2}~{\rm Hz},
\label{eq:c}
\end{equation}

where $z$ is the redshift of the source, $n$ the medium particle density, $E_{\rm k}$ the kinetic energy of the outflow, $t$ the time delay after the GRB and $C_p = 13(p-2)/\left[3(p-1)\right]$. We will assume for the micro-physical parameters $\epsilon_{\rm e}$, the fraction of total energy going to electrons, and  $\epsilon_{\rm B}$, the fraction of total energy going to magnetic fields, the values of $0.1$ and $0.01$, respectively. These figures are typical values measured during late-time afterglows \citep[see e.g.][]{PaKu02,Yos03} and in agreement with the results of the analysis of GRB\,080430. The relation for the cooling frequency is approximate since we are neglecting the possible role played by additional inverse Compton (IC) cooling. We will consider this issue again in Sect.\,\ref{sec:ssc}. 

The total energy can be derived from the burst isotropic energy E$_{\rm iso}$ with some assumptions about the spectrum and by correcting it for the fireball radiative efficiency $\eta$. We estimate E$_{\rm iso}$ as the integral of the burst spectral model \citep{Stam08} in the $1 - 10^4$\,keV band \citep{Ama02}, the energy range covering most of the prompt emission of GRBs. In this energy band the spectrum of a burst is typically described by a Band function \citep{Band93}:
\begin{eqnarray}
N_E(E) & = & \left\{\begin{array}{lccr}
A\left(\frac{E}{100keV}\right)^{\alpha} \mbox{exp}\left(-\frac{E}{E_0 }\right),
& ~ & (\alpha-\beta)E_0 \ge E \\
~ & ~ &  \\
A \left[\frac{(\alpha-\beta)E_0}{100 keV} \right]^{\alpha-\beta}
\mbox{exp}(\beta-\alpha) (\frac{E}{100keV})^{\beta}, & ~ &
(\alpha-\beta)E_0 \le E
\end{array}
\right.
\label{band_law}
\end{eqnarray}
In order to calculate the integral we need to know the two power-law photon indices $\alpha$ and $\beta$ and the peak energy E$_{\rm peak} = (2+\alpha)E_0$.  Unfortunately, the \textit{Swift}-BAT energy range is often too narrow for a direct E$_{\rm peak}$ measurement. 
We therefore run a set of integrations by varying E$_{\rm peak}$ and derive the corresponding E$_{\rm iso}$ using the $15 - 150$\,keV BAT fluence to normalize the spectrum. In each integration, depending on the value of E$_{\rm peak}$, we identify the observed photon index with one of the two indices of the Band function, fixing the other one to a canonical value (1 and 2.3 for the low- and the high-energy power-laws, respectively). The chosen values of E$_{\rm peak}$ and E$_{\rm iso}$ are those satisfying the Amati relation \citep{Ama02}. According to this method, we estimate E$_{\rm peak} = 39$\,keV, and E$_{\rm iso} = 3 \times 10^{51}$\,erg. The errors are estimated to be about 30\% for both quantities mainly due to the lack of observational data better constraining the prompt emission spectrum. Peak energies for \textit{Swift} GRBs can also be estimated following a correlation between peak energy and spectral parameters as measured by \textit{Swift}-BAT instrument \citep{Sak09} yielding consistent results with our analysis.
The derived total energy is typical for cosmological GRBs, although it is common to observe events substantially more energetic \citep{Sak08}. The relatively low estimated peak energy can allow one to classify this event as X-Ray Flash or X-Ray Rich \citep[][and references therein]{Zha07}. If we then assume a radiative efficiency $\eta$ of 10\%, we find the total kinetic energy going to the outflow E$_{\rm k, iso} = 3 \times 10^{52}$\,erg. The radiative efficiency during the prompt emission phase can vary among individual bursts \citep{Zhaet07}. A satisfactory treatment of the prompt emission phase emission process is still lacking. We choose 10\% as a conservative limit recalling it can be higher for events characterized by a shallow decay phase \citep{Nou06} in the X-rays as it might be the case for GRB\,080430.

Summing up, for modeling the high energy emission of the \object{GRB\,080430} afterglow, we have applied these parameters: energy  E$_{\rm k, iso} \sim 3 \times 10^{52}$\,erg, $\epsilon_{\rm e} \sim 0.1$, $\epsilon_{\rm B} \sim 0.01$, $p \sim 2.1$, the circumburst medium density profile $n \sim 1$\,cm$^{-3}$ and the redshift $z \sim 0.76$. Our observation was at $t \sim 8$\,ksec after the burst onset. At this epoch we have $\nu_{\rm m} \sim 2.1 \times 10^{12}$\,Hz and $\nu_{\rm c} \sim 8.8 \times 10^{15}$\,Hz. The afterglow synchrotron emission is in the so called "slow-cooling" regime (i.e. the synchrotron cooling frequency is above the synchrotron injection frequency) as confirmed by the modeling of the Spectral Energy Distribution (SED) from optical to X-rays (DEUG10) and usually expected at the epoch of the observations for typical afterglows \citep{ZhMe04}.

\section{Synchrotron-Self Compton during the afterglow}
\label{sec:ssc}

The analysis of the high-energy emission from the various phases of a GRB has been considered by many authors as a powerful diagnostic tool of GRB physics \citep{DeFr08,GaPi08,FaPi08,Pan08,Aha08,Falc08,Cov09,LeDe09,KuBD09,Fan09,Xue09,Gilm09,Mur09}. In the present case, the most important emission process to consider is essentially the Synchrotron-Self Compton (SSC).  Due to the long delay between the MAGIC observations and the GRB onset (about two hours) any residual prompt emission can be ruled out. Superposed to the SSC component, External Inverse Compton (EIC) processes could also play a role and will be briefly mentioned later. We do not consider here hadronic models \citep{BoDe98,PeWa05} in our discussion. They could, however, be of special interest if GRBs are important sources of cosmic-rays.

Once the parameters of the lower-energy synchrotron emission are known, it is possible to predict the SSC component with good reliability. Among the many possible choices, we followed the recipe described by \citet{FaPi08}.

The SSC process essentially generates a new spectral component superposed to the underlying synchrotron spectrum, with the same global shape up to a cut-off frequency:

\begin{equation}
\nu_{\rm M,SSC} \sim \frac{\Gamma^2 m_{\rm e}^2 c^4}{h^2 \nu_{\rm c}},
\end{equation}
where $\Gamma$ is the fireball bulk motion Lorentz factor, $m_{\rm e}$ the electron mass, $c$ the speed of light and $h$ the Planck constant. Above this frequency the SSC emission is no more in the Thomson regime and becomes much weaker (Klein-Nishina regime). For typical bulk motion Lorentz factors \citep[$\Gamma \sim 200$ at the afterglow onset,][]{Mol07} the SSC emission of the afterglow is in the Thomson regime.

Assuming we are in a constant density circumburst environment, the predicted SSC spectrum is characterized by two typical frequencies \citep{FaPi08} as the synchrotron afterglow spectrum (Sect.\,\ref{sec:aft}):

\begin{eqnarray}
\nu_{\rm m,SSC} \approx & 6.2 \times 10^{21}~C_p^4\,\epsilon_{{\rm e},-1}^4\,\epsilon_{{\rm B},-2}^{1/2}\,n^{-1/4}  \nonumber \\
& E_{\rm k,53}^{3/4}\,t_3^{-9/4}\,(1+z)^{5/4}~{\rm Hz},
\label{eq:mssc}
\end{eqnarray}

\begin{eqnarray}
\nu_{\rm c,SSC} \approx & 4 \times 10^{24}~(1+Y_{\rm SSC})^{-4}\,\epsilon_{{\rm B},-2}^{-7/2}\,n^{-9/4} \nonumber \\
& E_{\rm k,53}^{-5/4}\,t_3^{-1/4}\,(1+z)^{-3/4}~{\rm Hz},
\label{eq:vssc}
\end{eqnarray}
where $Y_{\rm SSC} = U'_{\rm syn}/U'_{\rm B}$ is the rest frame synchrotron to magnetic field energy density ratio. 

Defining $\xi_{\rm c}  = (\nu_{\rm m}/\nu_{\rm c})^{(p-2)/2}$, it can be shown \citep{SaEs01}  that\footnote{Here we deliberately ignore the possibility to have higher order IC components which could be effective in cooling the electron population.}:

\begin{equation}
Y_{\rm SSC} \simeq \frac{-1+\sqrt{1+4\xi_{\rm c}\epsilon_{\rm e}/\epsilon_{\rm B}}}{2}.
\end{equation}

The synchrotron injection to cooling synchrotron frequency ratio for the slow-cooling case is:

\begin{equation}
\nu_{m} / \nu_{c} \simeq 0.024\,(1+z)\,C_p^2\,\epsilon_{{\rm e},-1}^2\,\epsilon_{{\rm B},-2}^2\,n\,E_{\rm k,53}\,t_3^{-1}.
\label{eq:nmncr}
\end{equation}

The numerical factor in front of Eq.\,\ref{eq:nmncr} is not exactly the one derived from  Eq(s).\,\ref{eq:v} and \ref{eq:c} since, as already mentioned in Sect.\,\ref{sec:aft}, IC cooling also affects the location of the synchrotron cooling frequency making the problem numerically difficult to solve. We now apply an approximate solution fully adequate for our goals \citep[see][for a full discussion]{FaPi08}. With our parameters Eq.\,(\ref{eq:nmncr}) becomes $ \nu_{m} / \nu_{c} \simeq 0.00025 $ and $Y_{\rm SSC} \simeq 2.1$. Eqs.\,(\ref{eq:mssc}) and (\ref{eq:vssc}) become $\nu_{\rm m,SSC} \approx 1.1 \times 10^{18}$\,Hz ($\simeq 5$\,keV)  and $\nu_{\rm c,SSC} \approx 7.4 \times 10^{22}$\,Hz ($\simeq 310$\,MeV). The cooling SSC frequency is at much lower energy than the band covered by the MAGIC observations (E$_{\rm MAGIC} \sim 90$\,GeV). We are therefore in the spectral range where the SSC spectrum is softer, following a power-law behaviour, $\nu^{-p/2} \simeq \nu^{-1.05}$. 

In order to derive the expected flux density at the MAGIC energy we have to compute the flux density at the typical SSC frequency \citep{FaPi08} at the epoch of the MAGIC observation:

\begin{eqnarray}
F_{\nu_{\rm m},{\rm SSC}} \simeq & 7 \times 10^{-13}~n^{5/4} \epsilon_{{\rm B},-2}^{1/2}\,E_{\rm k,53}^{5/4}\,t_3^{1/4}\,\left(\frac{1+z}{2}\right)^{3/4} \nonumber \\
& D_{\rm L,28.34}^{-2}~{\rm erg\,cm}^{-2}\,{\rm s}^{-1}\,{\rm MeV}^{-1},
\end{eqnarray}
where $D_{\rm L}$ is the luminosity distance of the source, $D_{\rm L} \sim 4.8$\,Gpc ($\sim 1.5 \times 10^{28}$\,cm). With our parameters, $F_{\nu_{\rm m},{\rm SSC}} \simeq 5.2 \times 10^{-13}$\,erg\,cm$^{-2}$\,s$^{-1}$\,MeV$^{-1}$ which is much lower than the synchrotron flux at the same frequency, $\nu_{\rm m,SSC}$, well within the \textit{Swift}-XRT energy range with these parameters. 

Then, finally, from the peak energy to the MAGIC band we have to extrapolate the SSC spectrum as:

\begin{equation}
F_{90\,{\rm GeV}} \sim F_{\nu_{\rm m,SSC}} (\frac{\nu_{\rm c,SSC}}{\nu_{\rm m,SSC}})^{-(p-1)/2}\,(\frac{\nu}{\nu_{\rm c,SSC}})^{-p/2},
\label{eq:extrap}
\end{equation}
and, again with our parameters, $F_{90\,{\rm GeV}} \sim 2.9 \times 10^{-18}$\,erg\,cm$^{-2}$\,s$^{-1}$\,MeV$^{-1}$. The flux integrated in the MAGIC band, the parameter to be compared to the reported upper limits, can be well approximated by $\nu F_\nu$ at about 90\,GeV, and we have $F_{\rm MAGIC}  \sim 2.6 \times 10^{-13}$\,erg\,cm$^{-2}$\,s$^{-1}$ at the epoch of the MAGIC observation, $t \sim 8$\,ks from the burst.

Any uncertainty in the underlying afterglow parameters affects of course the VHE predictions. Some of these
uncertainties have, however, a rather limited (considering the present observational limits) impact because one of the relevant factors, the ratio between the injection and cooling synchrotron frequency, is constrained by the afterglow SED and uncertainties for micro-physical parameters should still keep the ratio close to the observed value. The $\nu_{\rm m}/\nu_{\rm c}$ ratio drives the importance of the IC component and the position of the cooling SSC frequency, i.e. where the VHE flux begins to decrease steeply moving toward higher energies. The total energy on the contrary is estimated assuming an efficiency for the GRB prompt emission process. This is a weakly known factor given that at present no satisfactory description of the GRB prompt emission process exists \citep{Lyut09}. It is therefore possible \citep{Zha07} that the efficiency is substantially higher, modifying the total energy and therefore the expected flux. Circumburst matter density has an important effect on the expected SSC flux. With the present afterglow data it can essentially only be estimated coupled to the micro-physical parameters. A higher density would make the SSC component more important and possibly detectable at lower energies \citep[see e.g.][]{Harr01}. However, the value of the circumburst density derived for afterglows with data allowing a detailed modeling is consistent with the value we report for GRB\,080430 \citep{Pan05}. 

A milder than expected temporal decay in the X-rays band together with the consistency of the observed SED with the reference afterglow model prediction, raises some concern about the reliability of the adopted theoretical scenario. A shallower afterglow decay showing a synchrotron spectrum can be explained with late-time energy injection in the outflow \citep{Pan06,Zha06}. In this case the VHE SSC flux temporal decay could be slowed in a way related to the time evolution of the energy injection \citep{WeFa07,GoMe07,GaPi07,FaPi08}. However, the lack of a similar behaviour at optical wavelengths do not fully support this possibility since energy injection should affect the afterglow evolution in any band. It could be possible that the X-ray afterglow is affected by the occurrence of a late-time slow flares which could be barely detectable at lower energies. Such a flare can produce a detectable VHE emission although likely with a longer and smoother time evolution due to the interaction of the flare photons with the outflow accelerated electrons \citep[see][]{FaPi08}, i.e. probably at later time than the MAGIC observations. Finally, we mention that micro-physical parameters can evolve in time. Their evolution could affect the position and time-evolution of the SSC injection and cooling frequencies and as consequence the expected VHE flux. However, a satisfactory theoretical framework for these possible modifications of the reference afterglow model is still lacking, leaving the introduction of these ingredients purely phenomenological and likely beyond the scope of this paper.

\begin{figure}[!t]
 \centering
 \includegraphics[width=\columnwidth]{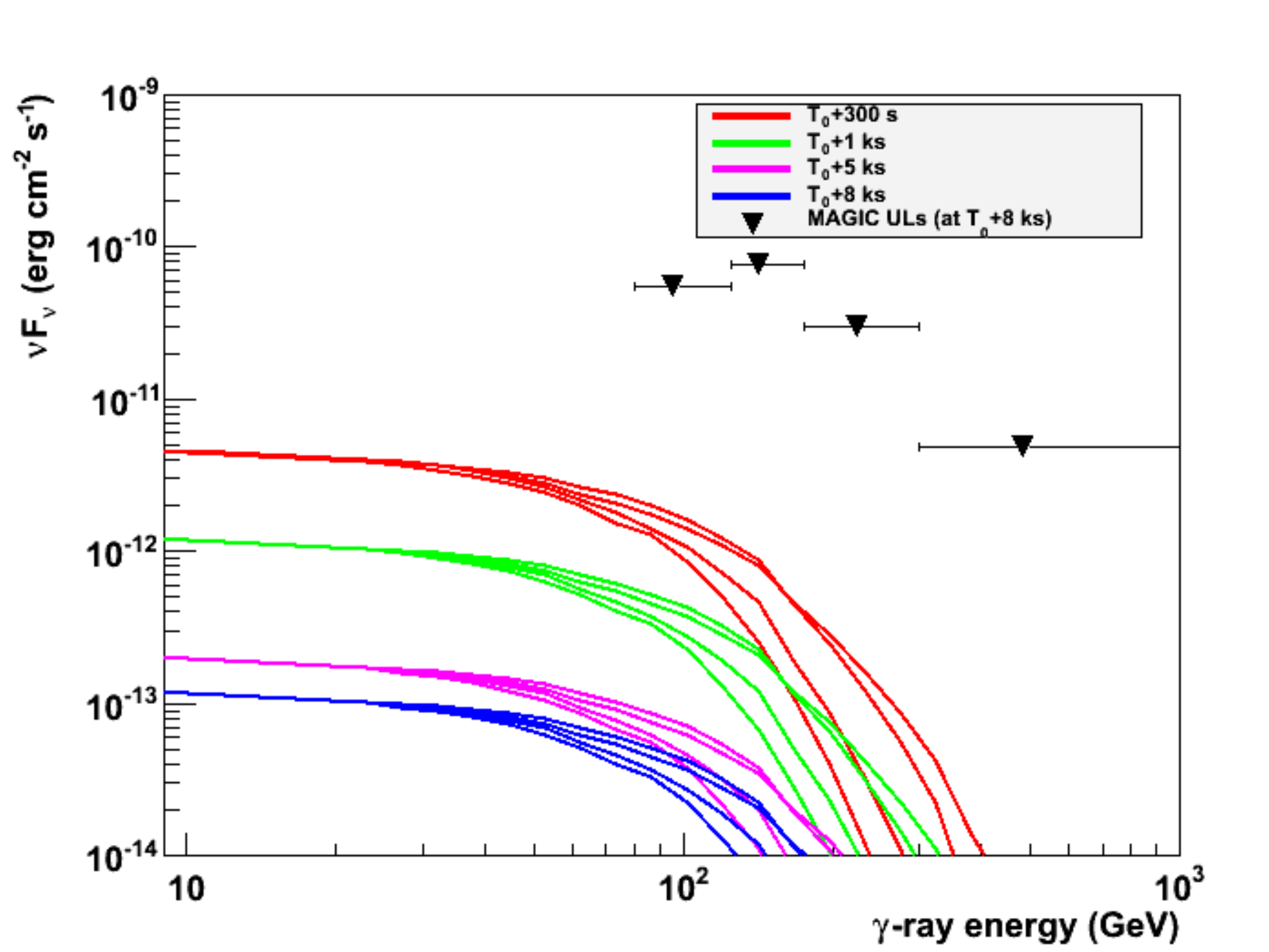}
 \caption{Predictions at different time delays from the high-energy event for the SSC emission during the afterglow of GRB\,080430. Black triangles are 95\% CL upper limits derived by MAGIC at various energies. Lines of a same color show the same SSC model, but a different absorption model of the gamma-rays by the EBL. The blue lines correspond to the MAGIC observation window.}
 \label{fig:ul}
\end{figure}

\section{Extragalactic background light attenuation}
\label{sec:ebl}

Gamma-rays in the GeV energy regime are absorbed through pair-production processes with the EBL. The precise light content of the EBL is strongly debated.  We have to rely on many different models, the predictions of which at $z \sim 1$ span a wide range of optical depths, from less than 1 up to 6 \citep{FaPi08}. Moreover, the MAGIC collaboration recently published a striking observational result \citep{Alb08} suggesting that the EBL attenuation could be much lower than previously assumed. Thus at the redshift of GRB\,080430 ($z \sim 0.76$) and at the MAGIC energy (E $\sim 90$\,GeV) an optical depth $\tau$ not far from unity is possible. We included four representative models from \cite{Knei04}, \citet{Fra08} and \citet{Gilm09EBL} and show the range of possible absorbed spectra in Fig.\,\ref{fig:ul}. The blue lines correspond to the MAGIC observation delay, the other lines show the spectrum at earlier observation times, in principle easily accessible to IACTs. On average, we can assume an attenuation of the received flux from the afterglow of GRB\,080430 of the order a factor 3 or even  less, allowing us to estimate $F_{\rm MAGIC}  \sim 3 \times 10^{-13}$\,erg\,cm$^{-2}$\,s$^{-1}$ as the predicted flux in the MAGIC band. As a matter of fact, our choice is possibly very conservative as  \citet{Gilm09EBL} described models, in agreement with the observations reported in \citet{Alb08}, with an optical depth as low as $\tau \sim 0.4$ at the same conditions of these MAGIC observations.

\section{Discussions}
\label{sec:concl}

The prediction of the expected SSC flux for an afterglow is not straightforward since it is required to know, or at least to reliably estimate, the parameters of the underlying afterglow (see Fig.\,\ref{fig:ul}). In the case of \object{GRB\,080430} the sampling of the X-ray and optical afterglow allowed us to estimate the various afterglow parameters to derive meaningful predictions for the expected SSC flux. However, a satisfactory modeling of the \object{GRB\,080430} can not be obtained within the standard fireball scenario. At least two different components are required for the early-time afterglow, as discussed in detail in DEUG10. Our present discussion is based on the assumption that one of these components is the regular afterglow \citep[i.e. the forward-shock][]{Pir99,Zha07} which is the main responsible for the late-afterglow emission although other components are likely playing a role. 

The results appear to be well below the reported upper limits. Furthermore, our assumed low opacity for the EBL is in agreement with current observations \citep[see also][]{Gilm09EBL}. At any rate, this pilot case shows fairly interesting perspectives for a late-afterglow detection at high energies. 

In general, to increase the flux expected from a GRB afterglow (for SSC) it is mandatory to try to decrease the observation energy (due to the $\nu^{-p/2}$ dependence above the cooling SSC frequency), which is also very important for the minimization of the EBL attenuation. If the telescope sum trigger hardware upgrade had already been implemented before the observations, a limit above an energy of 45\,GeV would have been obtained \citep[see also][]{GRB090102}. At these energies, the strong effect of the EBL could probably be neglected and the low energy threshold together with the expected performances of MAGIC\,II would undoubtedly increase the chances of positive detections. 

As a matter of fact, \object{GRB\,080430} was an average event in terms of energetics. More energetic GRBs are indeed relatively common, and due to the positive dependence on the isotropic energy of a GRB, much higher fluxes than in the present case can be foreseen. This is also true if we consider the uncertainty in the present total energy determination, which is based on an average value for the prompt emission efficiency.  

The time delay of the observation from the GRB has a clear impact, essentially because the observed SSC component is strictly related to the underlying synchrotron component which rapidly decays in intensity with time, depending on the specific environment and micro-physical parameters. Eq.~\ref{eq:extrap} goes roughly with $t^{-1.1}$ which means that had MAGIC been able to start observations right at the start of the late afterglow phase (e.g. at $T_0 + 1$ks), the flux predictions would have increased by more than an order of magnitude. 
The time delay of about two hours, coupled with the poor observing conditions, were more than enough to depress the observed flux and raise the reported upper limits. 

Given the uncertainties in the modeling of the afterglow, many possible modifications to the standard afterglow model \citep{Pir99,Zha07} can be applied. In some scenarios, substantially higher VHE flux can be predicted \citep[e.g.][]{Pan08,Mur10}, which makes observations at VHE energies powerful potential diagnostic tools.

The case of \object{GRB\,080430} in this pilot study demonstrates that if three conditions are met: 1) a moderate redshift ($z \lesssim 0.8$), 2) start of observations right at the beginning of the afterglow phase or even during the prompt emission and 3) the use of the MAGIC sum trigger enabling reaching energy thresholds below 50\,GeV, detection is within reach. The recent detection of $\sim 30$\,GeV photons during the prompt or afterglow phases of \object{GRB\,090510} \citep{Abd09} and \object{GRB\,090902B} \citep{deP09a,deP09b} by the Fermi satellite \citep{Band09} indeed shows that, with a threshold energy of a few tens of GeV and with the collecting area of a ground-based Cherenkov telescope, GRB VHE astrophysics is becoming a promising observational field.

\begin{acknowledgements}
We would like to thank the Instituto de Astrofisica de Canarias for the excellent working conditions at the Observatorio del Roque de los Muchachos in La Palma. The support of the German BMBF and MPG, the Italian INFN and Spanish MICINN is gratefully acknowledged. This work was also supported by ETH Research Grant TH 34/043, by the Polish MNiSzW Grant N N203 390834, and by the YIP of the Helmholtz Gemeinschaft. We also thank Yizhong Fan for continuous theoretical support. Lorenzo Amati, Cristiano Guidorzi, Alessandra Galli, Daniele Malesani and Ruben Salvaterra for useful discussions. We finally remark the very constructive report from the referee which helped to substantially improve the paper.
\end{acknowledgements}

\end{document}